\documentstyle[preprint,aps,amsfonts,amssymb,psfig]{revtex}
\begin{document}
\draft
\title{Vortex dynamics in disordered type-II superconductors}
\author{J. M\"ullers and A. Schmid}
\address{Institut f\"ur Theorie der Kondensierten Materie,
Universit\"at Karlsruhe, 76128 Karlsruhe, Germany}
\date{\today}
\maketitle
\begin{abstract}
A field theoretical method is developed which permits us to study the dynamics
of vortices in disordered environments. In particular,
we obtain a self-consistent system of equations for disorder averaged
quantities. Making use of a Hartree-type approximation, we calculate the
current-voltage (I-V) characteristics in the flux  flow regime.
In order to probe incipient melting of the vortex lattice
we propose an experiment where interference steps in the I-V characteristics
are observed, which arise when
a small ac field is superimposed on the constant voltage.
\end{abstract}
Superconductors of type II allow the magnetic field to enter in quantized
units of magnetic flux. Commonly one refers to the state with vortex lines
present as the Shubnikov phase and in this phase with an applied transport
current, an electric field may appear when the vortices are moving.

In an ideal situation,
the vortices are ordered in the form of a lattice. Material inhomogeneities
(static disorder \cite{LOJLTP}) tend to prevent
the free displacement of the vortices.
On the other hand, there are also thermal fluctuations (dynamic disorder)
which enhance the vortex motion. Thermal fluctuations are expected to
play an important role in high-temperature superconductors
\cite{Blatter}.

It is appropriate to distinguish two regimes for the current carrying Shubnikov
phase depending on whether the vortices are essentially at rest or in motion.
The first case is of interest with regard to
applications whereas in the second case, intrinsic properties of the vortices
and their dynamics will dominate the physics. In order to support the above
classification, we wish to draw attention to a recent experiment \cite{Yaron}
where by the help of neutron scattering the two regimes can be observed.

The theoretical analysis of these problems was mainly done considering
static situations\cite{Bou} or using perturbation theory for small fluctuations
\cite{CCreep}. In the field of research on disordered systems, the
techniques used for analyzing problems are
the replica method\cite{Bou} and supersymmetry\cite{Ef}. These
remove the so-called denominator problem in the partition
function (generating functional) when the average over the disorder is taken.
If one has to consider dynamical and interacting systems, the methods
referred to above do not seem to work well.

In this context, we recall a real-time method based
on the formalism of Martin, Siggia, and Rose (MSR)\cite{MSR,Phy} that permits
the calculation
of averaged dynamical quantities. It was applied by several authors, e.g.
in the fields of manifolds in random media\cite{Kinzel},
polymer melts\cite{Vil},
charge density waves\cite{NF}, and vortices in type-II superconductors
\cite{RFr}.

We present here a reformulation of the MSR theory, which is based on the
work of Cornwall, Jackiw, and Tomboulis (CJT)\cite{CJT,He}
and which leads to a self-consistent procedure that comprises an
expansion not only about the true values of the averaged fields,
but also about the true values of the correlation and response functions.
The main point of this method is a Legendre transform which yields the
velocity as an independent variable instead of the force. We think that
this choice is more suitable to describe the situation of an Abrikosov
lattice moving in the flux-flow regime. On the other hand, using a rather
direct Hartree approximation\cite{Kinzel}, it was shown that in the static
limit
ergodicity can be broken and aging effects may occur.

We apply the CJT method to analyze the dissipation due to
the motion of vortices in a disordered superconductor. Twenty years
ago, this problem was considered by one of us (A.S.)\cite{SH} as well
as by Larkin and Ovchinnikov\cite{LO2} in connection with
interference steps discovered by Fiory\cite{Fi2,Harris} in his experiments.
With the formalism presented below, the influence both of disorder and of
thermal noise on the vortex motion can be calculated in a systematic way.
As the high-$T_c$ superconductors are built of layers, there is a
fundamental interest in the behaviour of two-dimensional systems.
In addition, the above mentioned steps are more pronounced if superconducting
films are observed.
Therefore, as a starting point, we restricted ourselves to the analysis
of thin superconducting layers neglecting the vortex tilt.

Let us now consider the classical equation of motion for a two-dimensional
system of interacting particles $\{i\}$ (position $\bbox{r}^i(t)$, mass $m$,
harmonic interaction potential $V_{ij} \bbox{r}^i \bbox{r}^j/2$) moving
dissipatively (viscosity $\eta$) in a random potential $U(\bbox{r})$ and driven
by a force $\bbox{f}(t)$ as well as by Langevin forces $\bbox{\xi}^i(t)$.
For the rest of the article we shall use the following conventions: time
arguments will be written as index (e.g. $\bbox{r}^i_t$), thermal averages
are denoted by $\langle\cdot\rangle_T$, and
$\langle\cdot\rangle_U$ means the
average with respect to the random potential, which is taken to be Gaussian.
Thus, we have
\begin{eqnarray}
\label{motion}
        -\sum_j \int dt'\,{{\cal I}^R}^{ij}_{tt'} \, \bbox{r}^j_{t'} & := &
	m \ddot{\bbox{r}}_t^i + \eta \dot{\bbox{r}}_t^i + \sum_j
	V_{ij} \bbox{r}^j_t \nonumber \\
        & = & \bbox{f}_t - \nabla U (\bbox{r}_t^i) + \bbox{\xi}^i_t \,,
\end{eqnarray}
with according mean values and correlations $\langle U (\bbox{r})\rangle_U = 0$
and $\langle U (\bbox{r}) U (\bbox{r}') \rangle_U = \Gamma_U (\bbox{r}-
\bbox{r}')$ for disorder averages and $\langle \bbox{\xi}^i_t \rangle_T = 0$
as well as $\langle \bbox{\xi}^i_t \bbox{\xi}^j_{t'} \rangle_T =
2 \eta k_B T \delta^{ij}\delta_{tt'}$ for thermal averages.

The probability for a given path of the particles can be expressed by
the functional $\delta$-function, so that the generating functional for
the response and correlation functions is given by
\begin{eqnarray}
      Z ([\bbox{f}^i_t], [\bbox{J}^i_t]) & = &
      \int {\cal D}[\bbox{r}^i_t] {\cal D}[\bbox{R}^i_t]\,{\cal J} \nonumber\\
	& \times & \,e^{i \bbox{R} \bullet \left[ {{\cal I}^R}
       \bullet \bbox{r} + \bbox{f} + \bbox{\xi} -
       \nabla U(\bbox{r})
       \right]}\; e^{i \bbox{r} \bullet \bbox{J}} \,,
\end{eqnarray}
where e.g. $\bbox{r}\bullet\bbox{q}$ is meant to represent
$\sum_i \sum_\alpha \int dt\, r^{\alpha i}_t q^{\alpha i}_t$.
Since we have deliberately included the mass $m$ of the particles, the
Jacobian ${\cal J} = |\delta\bbox{\xi} / \delta\bbox{r}|$ is a
constant\cite{SchmQC} and may thus be omitted.

The above formulation, which is in essence the MSR theory,
allows to perform the disorder and thermal averages immediately.
After that, one can write the following compact expression for the
generating functional $Z$:
\begin{eqnarray}
\label{Zq}
     Z ([\bbox{\frak q}]) & = & {\cal J}\,\int {\cal D}[\bbox{\frak r}]\,
     e^{i \bbox{\frak q}\bullet\bbox{\frak r}}\;
     e^{i S([\bbox{\frak r}])} \,,  \\
     S([\bbox{\frak r}]) & = &
     (1/2) \bbox{\frak r}\bullet \bbox{\frak I}
        \bullet\bbox{\frak r}
     	- (i/2) \bbox{R}_\mu\bullet \partial^\mu \partial^\nu
        \Gamma_U (\bbox{r}-\bbox{r}') \bullet
        \bbox{R}'_\nu \nonumber
\end{eqnarray}
where we have introduced the symbols $\bbox{\frak r} =
   \left(\bbox{R},\bbox{r}\right)$,
$\bbox{\frak q} = \left(\bbox{f},\bbox{J}\right)$, and the matrix
\begin{equation}
	\bbox{\frak I} = \left( \begin{array}{cc} 2i\eta k_BT
	  & {\cal I}^R \\
	   {\cal I}^A & 0 \end{array} \right) \,;
\end{equation}
'$\bullet$' now includes an additional summation over the matrix indices
introduced above.
Note the structural similiarity between this matrix and the inverse
of a propagator in the Keldysh formalism\cite{Keld}.

In order to exploit the field theoretical formalism of Jackiw, Cornwall, and
Tomboulis (CJT)\cite{CJT} we proceed by adding a quadratic source term
 $(i/2) \bbox{\frak r}\bullet \bbox{\frak K} \bullet\bbox{\frak r}$
 to the exponent in (\ref{Zq}).
Taking the Legendre transform of the generating functional, that is, of the
vacuum diagrams $W ([\bbox{\frak q}], [\bbox{\frak K}]) = -i \ln Z$,
 we arrive at a set of equations which form the
starting point of the CJT formalism ($\overline{\bbox{\frak r}}$ means the
averaged position and $\bbox{\frak G}$ represents the connected correlation
and response functions of the particles):
\begin{eqnarray}
     \frac {\delta W}{\delta \bbox{\frak q}} & = &
        \overline{\bbox{\frak r}} \,,\qquad
     \frac {\delta W}{\delta \bbox{\frak K}} = \frac{1}{2} \left(
     	\overline{\bbox{\frak r}} \otimes \overline{\bbox{\frak r}} + i
     	 \bbox{\frak G} \right) \,, \\
     \Gamma ([\overline{\bbox{\frak r}}], [\bbox{\frak G}]) & = &
     	W ([\bbox{\frak q}], [\bbox{\frak K}]) - \bbox{\frak q}\bullet
     	\overline{\bbox{\frak r}} - \frac{1}{2} \overline{\bbox{\frak r}}
        \bullet \bbox{\frak K}
     	\bullet\overline{\bbox{\frak r}} - \frac{i}{2} \mbox{Tr}
     	\bbox{\frak G}\bullet \bbox{\frak K} \,;\nonumber \\
     \label{eqmot}
     \frac {\delta \Gamma}{\delta \overline{\bbox{\frak r}}} & = &
     	 - \bbox{\frak q}
     	-  \bbox{\frak K}\bullet\overline{\bbox{\frak r}} \,, \\
     \label{predyson}
    \frac {\delta \Gamma}{\delta \bbox{\frak G}} & = &
     	- \frac{i}{2} \bbox{\frak K} \,.
\end{eqnarray}
For the effective action $\Gamma$ one can write
\begin{eqnarray}
	\Gamma ([\overline{\bbox{\frak r}}], [\bbox{\frak G}])  & = &
	S([\overline{\bbox{\frak r}}]) - \frac{i}{2} \mbox{Tr}\ln{\bbox{\frak I}
	\bbox{\frak G}} + \frac{i}{2} \mbox{Tr} \frac{\delta^2
	S ([\overline{\bbox{\frak r}}])} {\delta \overline{\bbox{\frak r}}\,
	\delta \overline{\bbox{\frak r}}'}\,
	 \bbox{\frak G} \nonumber\\
	& & - \frac{i}{2} \mbox{Tr} \openone
	+ \Gamma_2 ([\overline{\bbox{\frak r}}],
	 [\bbox{\frak G}]) \,.
\end{eqnarray}
It has been proven by CJT to all orders, that $\Gamma_2$ can be represented as
\begin{equation}
\label{gamma2}
	\Gamma_2 ([\overline{\bbox{\frak r}}], [\bbox{\frak G}]) = \left.
	-i \ln{ \int {\cal D}[\bbox{\frak r}]\,
	e^{\frac{i}{2} \bbox{\frak r}\bullet\bbox{\frak G}^{-1}\bullet
	\bbox{\frak r} + i S_{int} ([\bbox{\frak r}],
	 [\overline{\bbox{\frak r}}])}} \right|_{2PI}
\end{equation}
where $S_{int}$ means an expansion of $S ([\bbox{\frak r} +
 \overline{\bbox{\frak r}}])$ about $\overline{\bbox{\frak r}}$ beginning
 from the third-order term in $\bbox{\frak r}$, and '2PI' indicates that only
 two-particle irreducible vacuum graphs in a theory with propagators
 given by $\bbox{\frak G}$ and vertices determined by $S_{int}$ are retained.

Note that eq. (\ref{predyson}) implies the Dyson equation
\begin{equation}
	\bbox{\frak G}^{-1} = \bbox{\frak K} + \tilde{\bbox{\frak I}}
	- 2 i \frac{\delta\Gamma_2}{\delta\bbox{\frak G}} =:
	\bbox{\frak K} + \tilde{\bbox{\frak I}} - \tilde{\Sigma} \,.
\end{equation}
If we set $\bbox{\frak K}=0$ and $\bbox{\frak q}=(\bbox{f},0)$, which is
the physical situation, the above Dyson equation together with the equation
of motion (\ref{eqmot})
constitues an exact self-consistent solution of the problem in consideration.

For further progress we resort to a Hartree-type approximation, which means
that in (\ref{gamma2}) the factor $\exp{(i S_{int})}$ is replaced by
$1 + i S_{int}$. This allows us to perform the path integral exactly because
only Gaussian integrals have to be evaluated.
In physical situations, $\overline{\bbox{R}}=0$, and $\bbox{\frak G}$
has Keldysh structure, i.e.
\begin{equation}
	\bbox{\frak G} = \left( \begin{array}{cc} 0 & {\cal G}^A \\
					{\cal G}^R & {\cal G}^K
			\end{array} \right) \,.
\end{equation}

For a demonstration
we apply the theory presented above to the simple system consisting of only
one particle moving in two dimensions subject to a random potential
("'tilted rough surface"') and to
Langevin forces. Specifically, the applied force $\bbox{f}_t$
is taken to be constant in time; as a consequence, the mean particle position
is $\overline{\bbox{r}} = \bbox{v} t$.
We may add that a constant mean velocity
requires a sufficiently large driving force, so that the probability for
having the particle trapped by a potential well is neglegible.
In order to complete the presentation of the simple model, we assume
that the random-potential correlator (in Fourier space) is
of the form $\Gamma_U (\bbox{k}) = \Gamma_U (0) \exp{(-k^2 \xi^2)}$.

Thus, in the overdamped limit the physical scales are determined by the
disorder strength $\Gamma_U(0)$, the viscosity $\eta$, as well as by the
correlation length $\xi$.
The self-consistent system of equation in this case reads
\begin{eqnarray}
	\label{selfeqs}
	{\cal G}^R_\omega & = & \frac{\hat{\bbox{v}}\otimes\hat{\bbox{v}}}
            {(m\omega+i\eta)(\omega+i0) - {\Sigma^R_L}_\omega} + \\
          & & + \frac{\openone - \hat{\bbox{v}}\otimes\hat{\bbox{v}}}
            {(m\omega+i\eta)(\omega+i0) - {\Sigma^R_T}_\omega}\,, \nonumber \\
	{\cal G}^K_\omega & = & {\cal G}^R_\omega {\cal G}^A_\omega\,
	   (\Sigma^K_\omega - 2 i \eta k_BT \openone) \,, \nonumber\\
	{\cal S}^R_t & = & \int \frac{d^2 k}{4 \pi^2} \, \Gamma_U (\bbox{k})
	   \bbox{k}\otimes\bbox{k}\,
	   e^{i \bbox{kv} t} e^{-i \bbox{k}({\cal G}^K_0 -
	   {\cal G}^K_t)\bbox{k}}\, \bbox{k} {\cal G}^R_t \bbox{k}\,,\nonumber\\
	\Sigma^R_\omega & = & {\cal S}^R_\omega - {\cal S}^R_0 \,, \nonumber\\
	\Sigma^K_t & = & - i \int \frac{d^2 k}{4 \pi^2} \, \Gamma_U (\bbox{k})
	   \bbox{k}\otimes\bbox{k}\,
	   e^{i \bbox{kv} t} e^{-i \bbox{k}({\cal G}^K_0 -
	   {\cal G}^K_t)\bbox{k}} \,, \nonumber\\
	\bbox{f}_P & = &
	   i \int \frac{d^2 k}{4 \pi^2} \,\Gamma_U (\bbox{k}) \bbox{k} \int dt\,
	   e^{i \bbox{kv} t} e^{-i \bbox{k}({\cal G}^K_0 -
	   {\cal G}^K_t)\bbox{k}} \,\bbox{k} {\cal G}^R_t \bbox{k} \,. \nonumber
\end{eqnarray}
The last line
represents the pinning force defined as $\bbox{f}_P := \bbox{f} -
\eta\bbox{v}$.
Note that we have $\bbox{f}\parallel\bbox{v}$ by symmetry.
The indices '$L$' and '$T$' denote the longitudinal and transverse parts
of the self-energy $\Sigma$,
 and we have used $(\bbox{a}\otimes\bbox{b})^{\mu\nu} := a^\mu b^\nu$.

\begin{figure}[p]
\vspace{2cm}
\vspace{1cm}
\caption{Pinning force $f_P (v; T)$ for a single particle (dimensionless units,
i.e. $\eta=1, \xi=1, \Gamma_U(0)=1$)}
\label{figFP}
\end{figure}
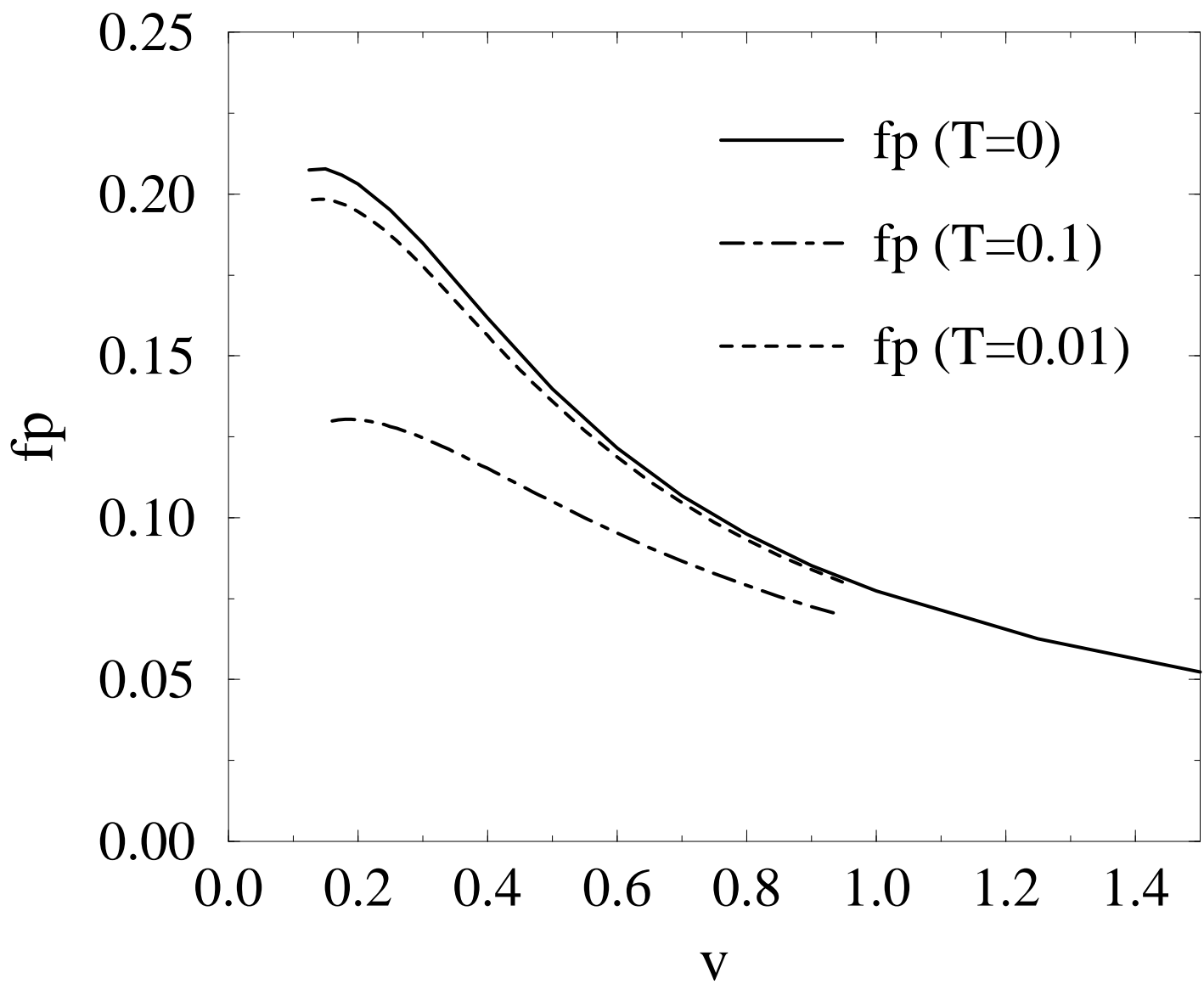

We have solved these equations numerically by iteration, i.e. starting by
rather arbitrary given self-energies and performing then repeatedly
the above steps. For the velocity $\eta v\gtrsim
\Gamma_U(0)^{1/2}/\xi^2$, the
iteration converges quite rapidly and yields
the force-velocity relation shown in fig. \ref{figFP}.
We have found that a diagonal approximation of all matrices, e.g.
${\cal G}^R \equiv G^R \openone$, leads to nearly the same curves.

To check the quality of the Hartee-type approximation, we have examined the
energy conservation. If we multiply the equation of motion (\ref{motion}) by
the particle velocity, average over random potential and thermal forces, and
subtract the unconnected part, we find
\begin{eqnarray}
\bbox{v}\,\bbox{f}_P & = &
\eta\, \langle\langle\, [\dot{\bbox{r}} - \bbox{v}]^2 \rangle\rangle -
\langle\langle \dot{\bbox{r}}\,\bbox{\xi} \rangle\rangle \nonumber\\
& \equiv &
-i\eta \ddot{G}^{K\alpha\alpha}_{t=0}
+ 2\eta k_BT \dot{G}^{R\alpha\alpha}_{t=0}\,.
\end{eqnarray}
The last line explicitly yields the relation
to the correlation and response functions.
Therefore, we can test our approximation by inserting the corresponding
functions calculated from eqs. (\ref{selfeqs})
into the conservation relation.
The agreement was found to be extremely well, and we believe that the
relation holds analytically even after the Hartree-type approximation.

For large velocities, $f_P$ and the self-energy $\tilde{\Sigma}$ vanish
 $\sim 1/v$. In this limit, a perturbation theory\cite{SH} is
 permissible.
On the other hand, for smaller velocities $\eta v \lesssim
\Gamma_U(0)^{1/2}/\xi^2$ the fluctuations grow
and finally become larger than the mean velocity. In this regime,
the importance of the self-energy $\tilde{\Sigma}$ comes into play.

After this preparatory discussion of a viscous particle moving on a rough
surface, we will now study the dynamics of vortex lattices.
The equation of motion (\ref{motion}) for this system can be cast in the form
\begin{equation}
	m \ddot{\bbox{r}}_{\bbox{l}t} + \eta \dot{\bbox{r}}_{\bbox{l}t}
	+ \sum_{\bbox{l'} \neq\bbox{l}}
	\bbox{D_{ll'}}\, \bbox{r_{l'}}_t = \bbox{f}_t
	- \nabla U (\bbox{l}+\bbox{r_l}_t) + \bbox{\xi_l}_t \,,
\end{equation}
where we have assumed harmonic vortex interactions\cite{Brandt}, and
$\bbox{l}$ denotes the equilibrium lattice sites. The dissipation is supposed
to originate from a Bardeen-Stephen mechanism \cite{BS}. In terms of
electrodynamical quantities the viscosity is related to the conductivity
by $\eta$ = $B \phi_0 \sigma_f$ where $\phi_0 = \pi\hbar/e$ is
the flux quantum. The velocity and force are related to
voltage and current by $\bbox{E}=\bbox{B}\times\bbox{v}$ and $\bbox{f}_L =
\bbox{j}_T\times\bbox{\phi}_0$ (Lorentz force).
The mass of the vortices is small and therefore, it may be omitted at the
last step of the calculations.

We shall analyze the flux-flow regime where for constant Lorentz force
$\bbox{f}$ the mean vortex positions are supposed to be $\bbox{l}+\bbox{v}t$.
One can go through the above sketched steps. In particular, we have done
numerical calculations for small lattices and have been able to confirm
the energy conservation. The expression used for the mean vortex position
implies that the vortices experience only
small deviations from the equilibrium
lattice positions. This assumption
should be reasonable for sufficiently large
velocities and is confirmed experimentally in a recent letter\cite{Yaron}.

Of special interest is a situation where in addition to the dc electric field
one has a small ac voltage of frequency $\Omega$\cite{Fi2,Harris},
so that we have
$\langle\langle \dot{\bbox{r}} \rangle\rangle
 = \bbox{v} + \bbox{w}\,\Omega \cos{\Omega t}$.
In that case, one can find steps in the $I_{dc}$-$V_{dc}$ curves at fields
$E_{nn'}=(n'/n) (\Omega/2\pi) \sqrt{2 \phi_0 B/\sqrt{3}}$.
This effect is based on the interference of the applied oscillations with
the motion of the {\it periodic} vortex lattice. It is completely analogous
to the mechanism responsible for the occurence of Shapiro steps\cite
{Shap2} in the I-V characteristics
of Josephson junctions. There, a mixing
of modes occurs due to the nonlinear (periodic) relation of phase and
current. For large junctions one finds interference steps for
$n' \hbar\Omega = n 2e V_{dc}$. This relation also has been pointed out
by Martinoli\cite{Mart}, who considered periodic pinning potentials.

In the case of disorder, the mixing of modes originates
from the nonlinearity of the random pinning potential. Thus,
one can gain information about the physics of the vortex lattice. From width
and height of the interference steps one can infer the shear modulus $c_{66}$
and the correlation $\Gamma_U (\bbox{r})$ of the pinning potential. This was
done by Fiory\cite{Fi2} using the perturbation theory of Schmid and
Hauger\cite{SH}. With our method, we may derive improved expressions
that are applicable at lower velocities.

For $4\pi \xi c_{66} \gg \eta v \gtrsim (\phi_0/4\pi B)^{1/2}
\Gamma_U(0)^{1/2}/\xi^2$ one has distinguishable steps and may
approximate the self-consistent scheme by the first nontrivial iteration.
In addition, we neglect thermal fluctuations and assume the ac amplitude
to be sufficiently small, so that one may restrict oneself to the consideration
of the steps $E_{n1}$ \cite{SH}; in an approximation, we take the matrix
$\tilde{\Sigma}$ from the dc case.

In order to have measurable effects of first order in $w$, one observes
the in-phase resistivity $\rho(E)$ \cite{Fi2}. Then, the height of the steps is
given by
\begin{equation}
   h_n = \frac{3 n^2 g_1^4 \rho_f^2}{32 c_{66} \Omega B \phi_0}\, \sum_m
   \Gamma_U (g_1 w_{mn}^{1/2}) w_{mn}
\end{equation}
with $w_{mn} = m^2 + mn + n^2$ and $g_1 = 2\pi (2 B/\sqrt{3}\phi_0)^{1/2}$.
If we expand $\rho(E)$ linearly near a step voltage where $E\approx E_{n1} +
\delta E_n/2$, we may define the halfwidth of the steps by
$\delta E_n = h_n / (2 \rho'(E_{n1}))$. Thus,
\begin{eqnarray}
   {(\delta E_n)}^{-1} & = & \frac{32\pi c_{66}\rho_f a_n n}
   {E_{n1}^2 \phi_0 g_1} \;
   \frac{\sum_m \Gamma_U (g_1 w_{mn}^{1/2}) w_{mn}}
     {\sum_{m'} \Gamma_U (g_1 w_{m'n}^{1/2}) w_{m'n}}\, \times \nonumber \\
   & & \times \left\{ 1-\sqrt{\pi} x_{mn} \exp{(x_{mn}^2)}\,
\mbox{erfc}(x_{mn})
    \right\}\,,    \nonumber \\
   a_n & := & 1 + \frac{2\pi g_1^2 \rho^2_f}{\sqrt{3} \phi_0^2 E_{n1}^2} \;
     \sum_{m'} \Gamma_U (g_1 m') {m'}^4\,, \nonumber\\
   x_{mn} & := & \frac{\pi^2 g_1^2 w_{mn} \rho_f^2 \sqrt{a_n}}
     {\sqrt{3\pi} \phi_0^2 E_{n1}^2} \;
     \sum_{m'} \Gamma_U (g_1 m') {m'}^2\,.
\end{eqnarray}

This result differs from the earlier one\cite{SH} by fluctuation effects which
appear here in a systematic way. For instance,
we obtain $i(G^K_{00}-G^K_{\bbox{l}-\bbox{l'},t-t'}) =
\langle\langle [\bbox{r}_{\bbox{l}t}-\bbox{r}_{\bbox{l'}t'} - \langle\langle
\bbox{r}_{\bbox{l}t}-\bbox{r}_{\bbox{l'}t'}\rangle\rangle]^2 \rangle\rangle
\propto \sqrt{t-t'}$ for large $|t-t'|$ and $\propto (t-t')^2$ for small time
differences, whereas in \cite{SH} a linear time dependence has been considered.
In addition, the contribution of $\Sigma^R$ is not included
in the analysis of \cite{SH}.

Note that the structure function is given by
\begin{equation}
	S(\bbox{k},\omega) = \sum_{\bbox{l}} \int dt\,
	e^{-i\omega t + i\bbox{kl} + i\bbox{kv}t}\,
	e^{-i k^2 (G^K_{00} - G^K_{\bbox{l}t})} \,;
\end{equation}
consequently, the broadening of the interference steps, which is governed by
the vortex fluctuations, is directly related to the structure factor. This
can be seen e.g. from eqs. (\ref{selfeqs}). Therefore, the width of the steps
can give indications for the beginning of a melting transition
\cite{Nel,Beck}
in the vortex lattice.

In summary,
we have presented a method that permits to study the dynamics of systems
subjected to random potentials and forces.
We have combined the formalisms of MSR and CJT and derived
a self-consistent system of equations for the averaged
particle positions and the correlation and response functions. The advantage
of this combination lies in the fact that it is not only an expansion
about the mean positions, as it would be the case for a mean field theory,
but also one about the full correlation and response functions.
Eventually, we have solved the self-consistent system of equations in a
Hartree type of aproximation. New results have been obtained for a single
particle moving on a rough surface in a viscous medium and also, for a
lattice of vortices.

We gratefully acknowledge the helpful
assistance of U. Eckern and H. Gl\"ockler.
This work was supported by the Deutsche Forschungsgemeinschaft.

\end{document}